\documentclass[12pt]{article}
\usepackage[paper=a4paper,margin=1in]{geometry}
\usepackage{t1enc}      
\usepackage{amsmath}
\usepackage{amsfonts}
\usepackage{amssymb}
\usepackage{amsthm}
\usepackage{graphicx}
\usepackage{mathrsfs}  

\newcommand{\ua}{\underline a \,}
\newcommand{\ub}{\underline b \,}
\newcommand{\uc}{\underline c \,}
\newcommand{\ud}{\underline d \,}

\newcommand{\nabbla}{{\nabla \!\!\!\!/}}

\numberwithin{equation}{section}

\begin{document}
\bibliographystyle{unsrt}

\title{On gravity's role in the genesis of rest masses of classical fields} 

\author{L\'aszl\'o B. Szabados\\
  Wigner Research Centre for Physics, \\
  H--1525 Budapest 114, P. O. Box 49, European Union}

\maketitle

\begin{abstract}
It is shown that in the Einstein--conformally coupled Higgs--Maxwell system 
with Friedman--Robertson--Walker symmetries the energy density of the Higgs 
field has stable local minimum only if the mean curvature of the $t={\rm 
const}$ hypersurfaces is \emph{less} than a finite critical value $\chi_c$, 
while for greater mean curvature the energy density is not bounded from 
below. Therefore, there are extreme gravitational situations in which 
\emph{even quasi-locally defined instantaneous vacuum states of the Higgs 
sector cannot exist}, and hence \emph{one cannot at all define the rest 
mass of all the classical fields}. On hypersurfaces with mean curvature 
less than $\chi_c$ the energy density has the `wine bottle' (rather than 
the familiar `Mexican hat') shape, and the gauge field can get rest mass 
via the Brout--Englert--Higgs mechanism. The spacelike hypersurface with the 
critical mean curvature represents the moment of `genesis' of rest masses. 
\end{abstract}


\section{Introduction}
\label{sec:1}

In the conformal cyclic cosmological (or CCC) model of Penrose \cite{PeCCC} 
all the particles and fields must be massless on a neighbourhood of the 
crossover hypersurfaces. (Such a hypersurface, as a regular spacelike 
hypersurface in the \emph{conformal spacetime}, is the interface between two 
successive aeons, and represents the future conformal boundary of the previous 
and the big bang singularity of the subsequent aeon.) Thus, according to this 
model, the particles and fields had to loose their rest mass in the very late 
stage of the previous aeon, and got rest mass only \emph{after} the Big Bang 
of our Universe in some mechanism. In particular, the particles and fields 
had to be massless in a neighbourhood of the Big Bang. Thus, in particle 
physics compatible with the CCC model, the rest masses should be expected to 
appear/disappear in some \emph{dynamical} process. 

The aim of the present paper is to clarify whether or not the rest mass of 
classical fields can appear/disappear in a dynamical process via the 
Brout--Englert--Higgs mechanism in a \emph{classical field theoretical} model 
which mimics all the characteristic feature of the Einstein--Standard Model 
system.

\subsection{The rest mass of relativistic classical fields}
\label{sub-1.1}

In the usual formulation of classical mechanics rest mass is an \emph{a 
priori given} attribute of particles, which can be determined from their 
small oscillations in a potential field around the \emph{stable equilibrium 
state} (see e.g. \cite{Ar}). Hence, in a more positivist approach to 
mechanics, the notion of rest mass can be \emph{introduced} via the study 
of small oscillations. 

In the relativistic theory of classical fields the \emph{definition} of the 
rest mass of the matter fields is based just on this idea. To illustrate 
this, let us consider a single real scalar field $\phi$ on the 
spacetime\footnote{The signature of the spacetime metric $g_{ab}$ is chosen 
to be $(+,-,-,-)$.} whose dynamics is governed by the Lagrangian ${\cal L}=
\frac{1}{2}g^{ab}(\nabla_a\phi)(\nabla_b\phi)-U(\phi)$, where the potential 
$U$ may depend on further fields, say $\Phi$, and it is a \emph{local, 
algebraic} expression of its variables\footnote{The potential $U$ is called 
a \emph{local, algebraic expression} of the field $\phi$ if its value $U
(\phi)(p)$ at any spacetime point $p$ is completely determined by the value 
$\phi(p)$ of the field there, and $U(\phi)(p)$ is an \emph{algebraic 
function}, e.g. a polynomial, of $\phi(p)$. Thus the derivatives of $U$ with 
respect to $\phi$ at the point $p$ are simply the derivatives of $U(\phi)(p)$ 
with respect to $\phi(p)$. These derivatives yield genuine smooth fields on 
$M$ rather than distributions.}. This yields the field equation $\nabla_a
\nabla^a\phi+(\partial U/\partial\phi)=0$. Following the mechanical analogy, 
if the scalar field is constant on the spacetime manifold $M$, say $\phi_0$, 
then, by $\nabla_a\phi_0=0$, this may be considered as a \emph{ground} or 
\emph{vacuum state} (being analogous to the \emph{equilibrium configuration} 
of the mechanical systems above). It solves the field equation precisely 
when $0=(\partial U/\partial\phi)_0=-(\partial{\cal L}/\partial\phi)_0$, 
where the subscript $0$ means `evaluated at $\phi_0$'. Thus, the ground 
states that solve the field equation are \emph{critical points} of the 
potential. Hence, if we write $U(\phi)=U(\phi_0)+(\partial U/\partial\phi)_0
(\phi-\phi_0)+\frac{1}{2}(\partial^2U/\partial\phi^2)_0(\phi-\phi_0)^2+...$, 
then the field equation yields 

\begin{equation}
\nabla_a\nabla^a\bigl(\phi-\phi_0\bigr)=-\Bigl(\frac{\partial^2U}
{\partial\phi^2}\Bigr)_0\bigl(\phi-\phi_0\bigr)+{\cal O}((\phi-\phi_0)^2). 
\label{eq:1.1}
\end{equation}
Let $p$ be any point of $M$ and $x^{\ua}$ the Gaussian normal (or local, 
approximate Cartesian) coordinate system\footnote{Latin indices from the 
beginning of the alphabet are abstract tensor indices, and the underlined 
indices are name indices, referring to some basis and taking numerical 
values, e.g. ${\ua}=0,...,3$.} on an open neighbourhood $W$ of $p$, which 
is based on an orthonormal vector basis $\{E^a_{\ua}\}$ and the origin at $p$ 
(see e.g. \cite{HE}). Let $k^a$ be a vector at $p$ with components $k^{\ua}$ 
in the basis $\{E^a_{\ua}\}$. Then (\ref{eq:1.1}) admits (local, approximate) 
solutions describing small linear oscillations on $W$ around $\phi_0$ with 
wave vector $k^a$, viz. $\phi_k(x^{\ua})=\phi_0+A\cos(k_{\ua}x^{\ua})+B\sin
(k_{\ua}x^{\ua})$ for $A,B\in\mathbb{R}$, precisely when 

\begin{equation}
g_{ab}k^ak^b=\Bigl(\frac{\partial^2U}{\partial\phi^2}\Bigr)_0(p). 
\label{eq:1.2}
\end{equation}
(Note that $(\partial^2U/\partial\phi^2)_0$ may still depend on the 
spacetime point $p$ via the other fields, say $\Phi$. Also, in this linear, 
approximate solution we neglect the back reaction of the scalar field to the 
spacetime geometry, i.e. this solution is considered to be only a `test 
field' to `scan' some of the local properties of the physical system.) 

By (\ref{eq:1.2}) the wave vector $k^a$ is spacelike, null or timelike 
precisely when the critical point $\phi_0$ of the potential is a maximum, an 
inflexion or a minimum point, respectively. Hence, the hypersurfaces in $W$ 
on which the solution $\phi_k(x^{\ua})$ has constant phase, $k_{\ua}x^{\ua}=
{\rm const}$, are, respectively, timelike, null or spacelike. Thus it is 
only the case of \emph{non-spacelike} wave vectors in which \emph{any} 
observer sees $\phi_k(x^{\ua})$ to be oscillating around $\phi_0$. For a 
spacelike wave vector there is a family of local Lorentz frames from which 
the solution $\phi_k(x^{\ua})$ appears to describe \emph{standing waves} 
rather than oscillations around $\phi_0$. The Lorentz invariant measure of 
the frequency of these oscillations is $g_{ab}k^ak^b$, i.e. by (\ref{eq:1.2}) 
it is given by the second derivative of the potential at the critical point. 

The \emph{standard definition} of (the square of) the rest mass $m$ of a 
field in classical field theory in \emph{flat spacetime} is just the second 
derivative of the potential with respect to the field at its critical 
points. In fact, it is \emph{precisely} this notion of rest mass that is 
used in the Brout--Englert--Higgs (BEH) mechanism \cite{H,EB} in the 
Standard Model of particle physics in Minkowski spacetime (see also 
\cite{AL}). (However, in traditional units, the physical dimension of this 
$m$ is 1/length rather than mass. Hence, in these units, the rest mass of 
the field $\phi$ is usually defined by $m^2:=(\hbar/c)^2(\partial^2 U/
\partial\phi^2)_0(p)$, even though $\phi$ is a \emph{classical} field, see 
e.g. \cite{HE,PR1,BD}.) Clearly, this rest mass is independent of the 
spacetime point $p$ precisely when there is a configuration $(\phi_0,
\Phi_0,...)$ in which all the fields are constant and which is a stable 
minimum of the potential. The usual vacuum states of Poincar\'e invariant 
field theories are typically such states (`spacetime vacuum state').

However, on general \emph{curved} spacetime this rest mass may depend on $p$ 
(but \emph{not} on the frame $\{E^a_{\ua}\}$); moreover the notion of rest 
mass could be introduced only \emph{quasi-locally}, on proper open subsets 
of $M$, like on $W$. An even more serious difficulty is when \emph{the 
potential $U$ does not have any minimum} with respect to $\phi$. In this 
case \emph{the rest mass of the field $\phi$ cannot be defined at all}. 
Then the field $\phi$ \emph{does not seem to have a particle interpretation} 
either. (We discuss some of the limitations of the particle interpretation 
of classical field configurations in the next subsection.) 

Since in the present paper primarily we are interested in how the BEH 
mechanism works in classical filed theoretical systems in extreme 
gravitational circumstances, in spite of these drawbacks and potential 
defects, we still adopt the above mathematical definition of the rest masses. 
We will see that there is, in fact, a physical system, viz. a $U(1)$ gauge 
field and a self-interacting complex Higgs field coupled to Einstein's 
general relativity in the conformally invariant way, in which the rest 
masses are not necessarily globally defined (i.e. can be introduced only 
quasi-locally) and do have a time dependence. In particular, we will see 
that in a neighbourhood of the initial singularity in a 
Friedman-Robertson-Walker spacetime \emph{the rest mass of the Higgs field 
is not defined at all} and the BEH mechanism does not work.

\subsection{On the particle interpretation  of relativistic classical 
field configurations on curved spacetime}
\label{sub-1.1*}

The very notion of rest mass is a genuine \emph{particle mechanical concept}, 
and it is not \emph{a priori} obvious that we can give a particle 
interpretation of the fields. In the present subsection we discuss the 
limitations of our ability to give such an interpretation in gravitational 
circumstances.\footnote{Thanks are due to one of the referees for suggesting 
to discuss the issues of this subsection in more details.} 

First, let us recall that the Gaussian normal coordinates $x^{\ua}$ are 
defined by the geodesics from $p$ in the direction $E^a_{\ua}x^{\ua}$ at $p$ 
\cite{HE}: The coordinates of a point $q\in M$ will be $x^{\ua}$ if $q$ is on 
the geodesic starting from $p$ with tangent $E^a_{\ua}x^{\ua}$ at $p$ and its 
affine distance from $p$ is 1. Thus, by the focussing effect of curvature on 
geodesics, the domain $W$ on which these coordinates can be introduced is 
typically only a \emph{proper subset} of $M$. The coordinates $x^{\ua}$ are 
not defined \emph{for, and beyond} the points where the neighbouring 
geodesics intersect each other. Hence, the approximate solution $\phi_k
(x^{\ua})$ is certainly not defined for arbitrarily large values of the 
coordinates. 

In addition, the coordinates $x^{\ua}$ are approximately Cartesian only on a 
much smaller neighbourhood of $p$: A straightforward calculation shows that, 
on a neighbourhood of $p$, the components of the metric and of the Christoffel 
symbols, respectively, are given by 

\begin{eqnarray*}
&{}&g_{\ua\ub}=\eta_{\ua\ub}+\frac{1}{6}\Bigl(R_{\ua\uc\ud\ub}+R_{\ub\uc\ud\ua}\Bigr)
 x^{\uc}x^{\ud}+{\cal O}(x^3), \\
&{}&\Gamma^{\ua}_{\ub\uc}=-\frac{1}{3}\Bigl(R^{\ua}{}_{\ub\uc\ud}+
 R^{\ua}{}_{\uc\ub\ud}\Bigr)x^{\ud}+{\cal O}(x^2);
\end{eqnarray*}
where $\eta_{\ua\ub}:={\rm diag}(1,-1,-1,-1)$, and $R_{\ua\ub\uc\ud}$ are the 
components of the curvature tensor in the basis $\{E^a_{\ua}\}$ at $p$. Thus, 
expanding the second derivative of the potential $U$ with respect to $\phi$ 
at the critical configuration in Taylor series around $p$, the field equation 
(\ref{eq:1.1}) in these coordinates takes the form 

\begin{eqnarray*}
&{}&\eta^{\ua\ub}\partial_{\ua}\partial_{\ub}(\phi-\phi_0)+\Bigl(\frac{\partial
  ^2U}{\partial\phi^2}\Bigr)_0(p)\bigl(\phi-\phi_0\bigr)= \\
&{}&-\frac{2}{3}x^{\ua}R_{\ua}{}^{\ub}\partial_{\ub}(\phi-\phi_0)-x^{\ua}\Bigl(
  \partial_{\ua}\Bigl(\frac{\partial^2U}{\partial\phi^2}\Bigr)_0\Bigr)(p)
  \bigl(\phi-\phi_0\bigr)+{\cal O}(x^2)+{\cal O}((\phi-\phi_0)^2). 
\end{eqnarray*}
Here $R_{\ua\ub}$ are the components of the Ricci tensor at $p$, and 
$\partial_{\ua}$ denotes partial derivative with respect to the coordinate 
$x^{\ua}$. 

Therefore, the root why $\phi_k(x^{\ua})$ fails to be an \emph{exact} 
solution of the field equation (\ref{eq:1.1}) is the non-vanishing of the 
terms on the right, e.g. the non-triviality of the spacetime curvature. In 
particular, the larger the Ricci curvature at $p$, the smaller the coordinate 
values for which $\phi_k(x^{\ua})$ is a good approximate solution. However, 
if the characteristic length of the spacetime curvature is much less than 
the wave length determined by (\ref{eq:1.2}), then the field equation may 
still have solutions, but these \emph{cannot be oscillating} solutions on a 
neighbourhood of $p$. Thus, in this case, the rest mass can still be well 
defined \emph{mathematically} by (\ref{eq:1.2}) even though its 
\emph{interpretation} as a `measure of inertia in small oscillations' is 
lost. In the main part of the present paper we will see that there could be 
situations in which not only the interpretation, but \emph{even the 
mathematical notion} of the rest mass above is also lost. Nevertheless, in 
the rest of the present subsection, we assume that $m^2=g_{ab}k^ak^b\geq0$ 
holds, $\phi_k(x^{\ua})$ exists on a neighbourhood of $p$, and we argue why 
this $m$ should be interpreted as the rest mass. 

The local, approximate solution $\phi_k(x^{\ua})$ determines an exact solution 
$\phi$ of (\ref{eq:1.1}) on $M$ such that, on (an open subset of) $W$, they 
coincide on a spacelike hypersurface $\Sigma$. In fact, since (\ref{eq:1.1}) 
is a second order hyperbolic partial differential equation, any of its 
solution $\phi$ (on a globally hyperbolic domain) is completely determined 
by its initial data set, consisting of its value $\psi$ and normal 
directional derivative $\dot\psi$, on a Cauchy surface $\Sigma$. Moreover, 
since this is not a constrained system, these two fields on $\Sigma$ can be 
chosen arbitrarily. In particular, on $\Sigma\cap W$, we can choose $\psi$ 
to be the value of $\phi_k(x^{\ua})$ and $\dot\psi$ to be its normal 
directional derivative; and then we can extend $\psi$ and $\dot\psi$ from 
$\Sigma\cap W$ to the whole of $\Sigma$ in an arbitrary, but smooth way. 
The corresponding exact solution $\phi$ of (\ref{eq:1.1}) coincides with 
$\phi_k(x^{\ua})$ on $\Sigma\cap W$ even in the first order in time by 
construction, and $\phi_k(x^{\ua})$ approximates $\phi$ on $W$ in the 
appropriate topologies discussed below. 

Clearly, this construction of initial data for the solutions of 
(\ref{eq:1.1}) works even if the wave vector $k^a$ in $\phi_k(x^{\ua})$ does 
not satisfy (\ref{eq:1.2}), e.g. when $p\in\Sigma$ and $k^a$ is tangent to 
$\Sigma$ even if the right hand side of (\ref{eq:1.2}) is strictly positive. 
In this case, the solution $\phi$ developing from this initial data is 
\emph{not} necessarily a propagating wave-like solution oscillating around 
$\phi_0$ on $W$, but it is e.g. a static one. In particular, we can consider 
the extensions of the initial data for $\phi_k(x^{\ua})$ such that outside a 
compact subset $K\subset\Sigma$ this is just the initial data set for 
$\phi_0$, i.e. $(\phi_0,0)$. Then, for all compact $K\subset\Sigma$ and any 
wave vector, they form a \emph{dense subset} in any open neighbourhood of 
$\phi_0$ even in the \emph{fine topology} (see e.g. \cite{Ha}). Clearly, the 
\emph{difference} of such a data set $(\psi,\dot\psi)$ and $(\phi_0,0)$ is 
finite in any $L_p$ norm, $1\leq p\leq\infty$. However, for non-compact 
$\Sigma$ and non-zero $\phi_0$, the $L_p$ norm of $\phi_0$ is finite only 
for $p=\infty$. Hence, the \emph{deviation} of this $(\psi,\dot\psi)$ from 
$(\phi_0,0)$ can be controlled in the $L_p$ (and hence in the $H^k_p$, $k\in
\mathbb{N}$, Sobolev) norms, even though $\phi_0$ itself does \emph{not} 
belong to any of these spaces (see e.g. \cite{AF}). 

The elementary local, linearized wave-like solutions $\phi_k(x^{\ua})$ (with 
wave vectors satisfying (\ref{eq:1.2})) provide not only a dense subset in 
neighbourhoods of $\phi_0$, but give justification why $g^{ab}k_ak_b$ should 
be interpreted as the (square of the) rest mass. To see this, let us recall 
that the energy-momentum tensor of the field $\phi$ is $T_{ab}=\nabla_{(a}
\phi\nabla_{b)}\phi-\frac{1}{2}g_{ab}(\nabla_c\phi)(\nabla^c\phi)+g_{ab}U$. 
Hence, in the leading order, for the energy and momentum densities of the 
perturbation $\phi_k(x^{\ua})-\phi_0$ of $\phi_0$, seen by the fleet of 
observers $(\partial/\partial x^0)^a$ on $W$, we obtain, respectively, that 

\begin{equation*}
\varepsilon-U(\phi_0)=\frac{1}{2}(\partial_0\phi_k)^2+\frac{1}{2}\delta^{ij}
(\partial_i\phi_k)(\partial_j\phi_k)+\frac{1}{2}m^2(\phi_k-\phi_0)^2, 
\hskip 20pt
\pi_i=(\partial_0\phi_k)(\partial_i\phi_k),
\end{equation*}
where $i,j=1,2,3$. Substituting the explicit form of $\phi_k(x^{\ua})$ here, 
for their \emph{average} in the coordinate 3-space $x^0={\rm const}$ on a 
box with length of edge $2\pi/\sqrt{\delta^{ij}k_ik_j}$ we obtain 

\begin{equation*}
{\tt e}=\frac{1}{2}\bigl(A^2+B^2\bigr)k_0^2, \hskip 20pt
{\tt p}_i=\frac{1}{2}\bigl(A^2+B^2\bigr)k_0k_i.
\end{equation*}
Hence, according to the special relativistic energy-momentum-rest mass 
relation, the `average rest mass' ${\tt M}$ of the perturbation $\phi_k
(x^{\ua})-\phi_0$ of $\phi_0$ in the 3-volume $V$ in the 3-space $x^0={\rm 
const}$ should be defined to be 

\begin{equation*}
{\tt M}:=\sqrt{{\tt e}^2-\delta^{ij}{\tt p}_i{\tt p}_j}V=\frac{1}{2}\bigl(
A^2+B^2\bigr)k_0V\,m.
\end{equation*}
As it could be expected, this ${\tt M}$ depends on the amplitudes $A$ and 
$B$, the frequency $k_0$ (measured in the $x^0={\rm const}$ 3-space in which 
the average was taken) and the volume $V$; but also it is proportional to 
$m$, defined by (\ref{eq:1.2}). Thus, it is the rest mass $m$ that is the 
\emph{common} property of \emph{all the linear wave-like perturbations} of 
the critical configuration $\phi_0$ near $p\in M$, and hence a property of 
the \emph{physical system} defined by the Lagrangian ${\cal L}$ and of the 
point $p$. (This result shows that $m$ can also be interpreted as the rest 
mass of the elementary Fourier modes, i.e. \emph{one-particle classical 
states}, rather than the rest mass of the field $\phi$ in general.) 

Finally, it might be worth noting that an independent justification of the 
interpretation of $m$ as the rest mass (of the one-particle states) is 
provided by an elementary quantum mechanical argumentation. Indeed, in the 
local Lorentz frame (using the traditional notations and units) $g_{ab}k^ak^b
=(\omega/c)^2-\delta_{ij}k^ik^j$. Then by the Planck--de Broglie hypothesis 
of elementary quantum theory the energy and linear momentum corresponding 
to the wave with frequency $\omega$ and spatial wave vector $k_i$, 
respectively, are $E=\hbar\omega$ and $p_i=\hbar k_i$; by means of which 
$g_{ab}k^ak^b=(E^2-c^2\delta_{ij}p^ip^j)/(c\hbar)^2$. Hence, if we want to 
keep the special relativistic energy-momentum-rest mass relation to be valid, 
then this should be identified with $(mc/\hbar)^2$, i.e. $m^2=(\hbar/c)^2
(\partial^2 U/\partial\phi^2)_0(p)$, just as we claimed in subsection 
\ref{sub-1.1}.

\subsection{The question of origin of the rest masses of relativistic 
fields}
\label{sub-1.2}

In field theory it is usually assumed that the system admits a vacuum (or 
ground) state, thus the rest mass of the fields is usually assumed to be 
well defined. This rest mass can be zero or non-zero, and there is an idea 
(according e.g. to the standard model, in particular the Weinberg--Salam 
model \cite{AL}, or to the twistor programme \cite{PeMc}) that fundamentally 
the fields have zero rest mass, and the origin of their non-zero rest mass 
is due to their interactions. This idea is formulated mathematically by the 
BEH mechanism: In the Weinberg--Salam model, the rest mass of the spinor and 
gauge fields is due to their interaction with the Higgs field, while that of 
the Higgs field to its own self-interaction. The resulting rest masses are 
proportional to the value of the Higgs field in its \emph{gauge symmetry 
breaking} vacuum state (see \cite{H,AL}). 

According to the standard view in astro-particle physics, in the very early 
era in the history of the Universe the Higgs field had only a symmetric  
vacuum state. Hence, the spinor and gauge fields could not get rest mass, 
and it was only a later stage when the symmetry breaking vacuum states 
emerged and the BEH mechanism started to work. Thus, according to this view, 
initially the fields \emph{did} have a vacuum state, which was 
\emph{symmetric}, i.e. invariant with respect to the gauge transformations. 
Hence, initially, all the fields had \emph{zero rest mass}, and they got 
non-zero rest mass later. If, however, \emph{no vacuum states} of the Higgs 
sector, neither symmetric nor symmetry breaking, existed at an early stage, 
then \emph{even the notion} of rest mass of the Higgs field, zero or 
non-zero, could not be introduced. Hence, we should clarify whether or not 
there are extreme circumstances in which such states do not exist, and we 
should find the \emph{mathematical criteria} of their existence. 

In the present note we are interested in the effect of gravitation in the 
BEH mechanism, when there is a direct conformally invariant coupling of the 
Higgs field to the gravitational `field', too. This coupling improves the 
conformal properties of the Higgs sector significantly (and the improved 
conformal properties could be considered as a mathematical realization of 
the idea that fundamentally the matter fields are massless, see e.g. 
\cite{PeMc}), but it does not yield any observable change in the low energy 
predictions of the Standard Model. For the sake of simplicity, the Higgs 
field will be a single complex, self-interacting scalar field $\Phi$, and 
the gauge field is a single gauge field $\omega_a$ with gauge group $U(1)$. 
(For the sake of simplicity, we call $\omega_a$ a Maxwell field, although 
it is not intended to describe electromagnetism.) We call this model the 
`Einstein-conformally coupled Higgs--Maxwell' (or shortly EccHM) system. 
Also to simplify the calculations of the vacuum states and the rest masses, 
we assume that the spacetime can be foliated by spacelike hypersurfaces 
with constant mean curvature and the mean curvature could be used as a time 
coordinate (`York's time'), like in the Friedman--Robertson--Walker (FRW) 
spacetimes. (A more realistic model with a gauge field with arbitrary 
compact gauge group and arbitrary Higgs and Weyl spinor multiplets is 
considered in \cite{Sz16}. There the analogous calculations for the 
Kantowski--Sachs spacetimes, e.g. for the metric inside a spherical black 
hole, are also given.) 

First we show that the `obvious' candidate for the global spacetime vacuum 
state of the EccHM system, represented by a spacetime with maximal Killing 
symmetry and a constant Higgs field minimizing the energy density and 
solving the field equation, does \emph{not} exist. Then we search for a 
weaker notion of vacuum states, the `instantaneous' ones on spacelike 
hypersurfaces: These are defined to be those states in which the matter 
fields admit the isometries of the spacetime as symmetries, solve the 
\emph{constraint} (rather than all the field) equations and minimize the 
energy \emph{functional}\footnote{Analogous instantaneous vacuum states in 
the quantum theory of linear scalar fields in FRW spacetimes have been 
introduced recently in \cite{ANA}.}. In the presence of FRW symmetries we 
determine the criteria of the existence and properties of these states: On a 
hypersurface $\Sigma$ instantaneous vacuum states exist only when the mean 
curvature $\chi$ of $\Sigma$ is less than a large, but finite critical 
value $\chi_c$; and when they exist, then they are necessarily gauge symmetry 
breaking and depend on $\chi$. The hypersurface on which $\chi=\chi_c$ is 
the `instant of the genesis' of the rest masses. The rest masses calculated 
via the BEH mechanism are \emph{time dependent}, and decreasing with 
decreasing $\chi$. However, their time dependence is essential only in a 
rather short period after their `genesis', and the predictions of the model 
reduce rapidly to the familiar results known from the flat spacetime field 
theory. 

In section \ref{sec:2} we specify the model, derive the field equations and 
the energy-momentum tensor; and show that the usual `natural candidate' for 
the global spacetime vacuum states does not exist. Then, in section 
\ref{sec:3}, we define the instantaneous vacuum states for the classical 
fields, and determine the criteria of their existence in FRW spacetimes and 
clarify their time dependence. Finally, in section \ref{sec:4}, we calculate 
the rest mass of the Higgs and the gauge fields via the BEH mechanism. 

Our sign conventions are those of \cite{Sz16}: In particular, Einstein's 
field equations take the form $R_{ab}-\frac{1}{2}Rg_{ab}=-\kappa T_{ab}-
\Lambda g_{ab}$, where $\Lambda$ is the cosmological constant and $\kappa:=8
\pi G$ with Newton's gravitational constant $G$. (In the $\hbar=c=1$ units 
the numerical value of these constants is given by $\Lambda=10^{-58}cm^{-2}$ 
and $6/\kappa=8.6\times10^{64}cm^{-2}$.)


\section{The Einstein-conformally coupled Higgs-Maxwell 
system}
\label{sec:2}

\subsection{The field equations and the energy-momentum tensor}
\label{sub-2.1}

Our basic matter field variables are the complex scalar $\Phi$ and the 
$U(1)$ gauge field $\omega_a$, whose dynamics are governed by the Lagrangian 

\begin{equation}
{\cal L}:=-\frac{1}{4}F_{ab}F_{cd}g^{ac}g^{bd}+\frac{1}{2}g^{ab}(\nabbla_a\Phi)
(\nabbla_b\bar\Phi)-\frac{1}{2}\mu^2\vert\Phi\vert^2-\frac{1}{4}\lambda
\vert\Phi\vert^4-\frac{1}{12}R\vert\Phi\vert^2, \label{eq:2.1}
\end{equation}
where $F_{ab}:=\nabla_a\omega_b-\nabla_b\omega_a$, the field strength of the 
gauge field, $\nabbla_a\Phi:=\nabla_a\Phi+{\rm i}\omega_a\Phi$ the 
gauge-covariant derivative of the scalar field, $R$ is the curvature scalar 
of the spacetime, and $\mu^2$ and $\lambda$ are real constants. (In the 
$\hbar=c=1$ units the parameters of the Standard Model are $\lambda=1/8$ 
and $\mu^2=-1.8\times10^{31} cm^{-2}$.) The corresponding field equations are 

\begin{eqnarray}
\nabla^aF_{ab}\!\!\!\!&=\!\!\!\!&\frac{\rm i}{2}\bigl(\bar\Phi\nabbla_b
  \Phi-\Phi\nabbla_b\bar\Phi\bigr)=:4\pi J_b, \label{eq:2.2a} \\
\nabbla_a\nabbla^a\Phi\!\!\!\!&=\!\!\!\!&-\bigl(\mu^2+\frac{1}{6}R\bigr)
  \Phi-\lambda\bar\Phi\Phi^2; \label{eq:2.2b}
\end{eqnarray}
while the energy-momentum tensor, defined to be twice the variational 
derivative of the matter action with respect to $g^{ab}$, is 

\begin{eqnarray}
T_{ab}\!\!\!\!&=\!\!\!\!&-F_{ac}F_{bd}g^{cd}+\frac{1}{4}g_{ab}F_{cd}F^{cd}+
  (\nabbla_{(a}\Phi)(\nabbla_{b)}\bar\Phi)-\frac{1}{2}g_{ab}(\nabbla_c\Phi)
  (\nabbla^c\bar\Phi)+\label{eq:2.3} \\
\!\!\!\!&+\!\!\!\!&\frac{1}{2}g_{ab}\mu^2\vert\Phi\vert^2+\frac{1}{4}g_{ab}
  \lambda\vert\Phi\vert^4-\frac{1}{6}\bigl(R_{ab}-\frac{1}{2}Rg_{ab}\bigr)
  \vert\Phi\vert^2-\frac{1}{6}\nabla_a\nabla_b\vert\Phi\vert^2+\frac{1}{6}
  g_{ab}\nabla_c\nabla^c\vert\Phi\vert^2; \nonumber
\end{eqnarray}
which is compatible with that for the conformally invariant scalar field 
given in \cite{NP68}. Its trace is $T_{ab}g^{ab}=\mu^2\vert\Phi\vert^2$, and 
hence, by Einstein's equations, the field equation for the Higgs field can 
be rewritten in the form 

\begin{equation}
\nabbla_a\nabbla^a\Phi=-\bigl(\mu^2+\frac{2}{3}\Lambda\bigr)\Phi-\bigl(
\lambda+\frac{1}{6}\kappa\mu^2\bigr)\bar\Phi\Phi^2. \label{eq:2.4}
\end{equation}
Note that its structure is just that of (\ref{eq:2.2b}) in flat spacetime: 
It is only the rest mass and self-interaction parameters that are shifted 
by $2\Lambda/3$ and $\kappa\mu^2/6$, respectively. Thus, on a \emph{given} 
spacetime, the \emph{structure} of the solutions of the field equations 
(\ref{eq:2.2a}), (\ref{eq:2.4}) is the same that of the Maxwell--Higgs system 
without the conformal coupling to gravity. On the other hand, if, using 
Einstein's equations, we substitute the Einstein tensor on the right hand 
side of (\ref{eq:2.3}) by the energy-momentum tensor and the cosmological 
constant, then even the \emph{structure} of the energy-momentum tensor 
changes significantly: 

\begin{eqnarray}
T_{ab}\!\!\!\!&=\!\!\!\!&\bigl(1-\frac{1}{6}\kappa\vert\Phi\vert^2\bigr)^{-1}
  \Bigl\{-F_{ac}F_{bd}g^{cd}+\frac{1}{4}g_{ab}F_{cd}F^{cd}+(\nabbla_{(a}\Phi)
  (\nabbla_{b)}\bar\Phi)-\frac{1}{2}g_{ab}(\nabbla_c\Phi)(\nabbla^c\bar\Phi)-
\nonumber \\
\!\!\!\!&-\!\!\!\!&\frac{1}{6}\nabla_a\nabla_b\vert\Phi\vert^2+\frac{1}{6}
  g_{ab}\nabla_c\nabla^c\vert\Phi\vert^2+\frac{1}{2}g_{ab}(\mu^2+\frac{1}{3}
  \Lambda)\vert\Phi\vert^2+\frac{1}{4}g_{ab}\lambda\vert\Phi\vert^4\Bigr\}. 
  \label{eq:2.5}
\end{eqnarray}
Thus the energy-momentum tensor, and hence via Einstein's equations the 
spacetime, may have \emph{two} different kinds of singularities: The first 
is when the matter field variables are diverging, and the second is when the 
pointwise norm of the Higgs field takes the special value $\vert\Phi\vert^2
=6/\kappa$. Since by Einstein's equations $R=4\Lambda+\kappa\mu^2\vert\Phi
\vert^2$, in the former case the curvature scalar is diverging if $\vert\Phi
\vert$ is diverging, but in the latter $R$ remains bounded. Thus, the second 
singularity is less violent than the first, and hence (motivated by the 
terminology `Big Bang' for the first in the cosmological context), we call 
the second the `Small Bang'. (A more detailed discussion of the general 
properties of these singularities, even in the general Einstein--conformally 
coupled Standard Model (EccSM) system, see \cite{Sz16}.) However, the 
configuration $\vert\Phi\vert^2=6/\kappa$ is not necessarily singular: The 
field equations of the Einstein-conformally coupled Higgs (EccH) system in 
the presence of FRW symmetries have solutions in which $\vert\Phi\vert^2=6/
\kappa$ corresponds to scalar polynomial curvature singularities of the 
spacetime, but there are solutions in which $\vert\Phi\vert^2=6/\kappa$ at 
regular spacetime points. (A more detailed discussion of these solutions 
will be published in a separate paper \cite{SzW}.)

\subsection{Non-existence of global spacetime vacuum states}
\label{sub-2.2}

In the lack of any well defined energy density of the gravitational `field', 
the usual definition of the vacuum states of classical field theories in 
Minkowski spacetime cannot be applied directly to the present case. However, 
on physical grounds it seems plausible to \emph{postulate} that the `vacuum 
states' of the EccHM system are represented by those matter+gravity 
configurations in which the spacetime is of maximal symmetry, and the matter 
fields admit these isometries as symmetries, solve the field equations and 
minimize the energy density. Hence the spacetime is of constant curvature 
(de Sitter, Minkowski or anti-de Sitter). Thus $R_{ab}-\frac{1}{2}Rg_{ab}=-
\frac{1}{4}Rg_{ab}$ holds, and the matter field variables are such that 
$F_{ab}=0$ and $\nabbla_a\Phi=0$. If, in addition, we assume that the $U(1)$ 
bundle of gauge field configurations is globally trivializable, then the 
gauge field can be chosen to be vanishing even globally on $M$. Hence, the 
Higgs field is constant on $M$, $\nabla_a\Phi=0$. These special 
matter+gravity configurations are analogous to the `equilibrium 
configurations' of subsection \ref{sub-1.1} of the introduction. 

Substituting $\omega_a=0$ and $\nabla_a\Phi=0$ into (\ref{eq:2.5}) we find 
that the energy-momentum tensor is a pure trace: 

\begin{equation}
T_{ab}=\frac{1}{2}\frac{\vert\Phi\vert^2}{1-\frac{1}{6}\kappa\vert\Phi\vert^2}
\Bigl(\mu^2+\frac{\Lambda}{3}+\frac{1}{2}\lambda\vert\Phi\vert^2\Bigr)g_{ab}
=:\frac{1}{4}Tg_{ab}. \label{eq:2.6}
\end{equation}
Thus, the energy density, seen by any local observer $t^a$, is $\varepsilon
:=T_{ab}t^at^b=\frac{1}{4}T$. Clearly, the field configurations $\omega_a=0$, 
$\nabla_a\Phi=0$ solve (\ref{eq:2.2a}), but (\ref{eq:2.4}) is not satisfied 
identically. It fixes the value of the norm of the Higgs field to be 

\begin{equation}
\vert\Phi_g\vert^2:=-\frac{\mu^2+\frac{2}{3}\Lambda}{\lambda+\frac{1}{6}
\kappa\mu^2}, \label{eq:2.7}
\end{equation}
while Einstein's equations yield that the spacetime is necessarily anti-de 
Sitter. However, $\Phi_g$ (`spacetime ground states') do \emph{not} minimize 
the energy density. In fact, the critical points of $\varepsilon$ are at 
$\Phi=0$ and at the solutions of 

\begin{equation}
-\frac{1}{12}\kappa\lambda\vert\Phi\vert^4+\lambda\vert\Phi\vert^2+\mu^2+
\frac{\Lambda}{3}=0. \label{eq:2.8}
\end{equation}
For the latter we obtain that 

\begin{equation}
\vert\Phi_\pm\vert^2=\frac{6}{\kappa}\Bigl(1\pm\sqrt{1+\frac{\kappa}{3\lambda}
\bigl(\mu^2+\frac{1}{3}\Lambda\bigr)}\Bigr). \label{eq:2.9}
\end{equation}
$\Phi=0$ and $\Phi_+$ are local \emph{maxima}, while the configurations 
$\Phi_v:=\Phi_-$ (`spacetime vacuum states') are the local \emph{minima} of 
$\varepsilon$. Comparing $\vert\Phi_g\vert^2$ and $\vert\Phi_v\vert^2$ we 
find that these two would coincide precisely when $\Lambda=-3\mu^2-9\lambda/
\kappa$ or $\Lambda=\kappa\mu^4/4\lambda$ held. Neither of these conditions 
is satisfied with the known numerical value of the constants $\kappa$, 
$\Lambda$, $\mu^2$ and $\lambda$ of the Einstein--conformally coupled 
Standard Model system. 

Therefore, the conformal coupling of the matter sector to gravity yields 
that the two key properties of the usual spacetime vacuum/ground states, 
viz. that they solve the field equations and minimize the energy density, 
split. Hence, the criteria in the notion of vacuum states should be weakened. 
This leads us to the concept of the `instantaneous vacuum states'. This is 
based on the 3+1 decomposition of the spacetime and will be defined and 
discussed in subsection \ref{sub-3.3} below.

\subsection{The 3+1 decomposition}
\label{sub-2.3}

Let the foliation of the spacetime by spacelike hypersurfaces $\Sigma_t$ be 
fixed, let $t^a$ be its future directed timelike normal, $N$ its lapse 
function, and let us choose an evolution vector field $\xi^a=Nt^a+N^a$, where 
the shift vector, $N^a$, is tangent to the leaves $\Sigma_t$. (For the basic 
notions in the 3+1 decomposition, see e.g. \cite{HE,IsNe}.) If $P^a_b:=
\delta^a_b-t^at_b$, the orthogonal projection to the leaves, then the induced 
metric on $\Sigma_t$ is defined by $h_{ab}:=P^c_aP^d_bg_{cd}$, and the extrinsic 
curvature of $\Sigma_t$ is $\chi_{ab}:=P^c_aP^d_b\nabla_{(c}t_{d)}$. Then the 
spacetime volume element is ${\rm d}v=N{\rm d}\Sigma{\rm d}t$. We decompose 
the gauge field and the field strength according to the conventions $\phi:=
\omega_at^a$, $A_a:=P^b_a\omega_b$, $E_a:=F_{ab}t^b$ and $B_{ab}:=P^c_aP^d_b
F_{cd}$. The time derivative of a purely spatial tensor field, say 
$S_{ab\cdots}=P^c_aP^d_b\cdots S_{cd\cdots}$, is defined by $\dot S_{ab\cdots}:=
P^c_aP^d_b\cdots({\pounds}_\xi S_{cd\cdots})$, where ${\pounds}_\xi$ denotes Lie 
derivative along the vector field $\xi^a$. In particular, if $D_a$ denotes 
the intrinsic Levi-Civita covariant derivative on $\Sigma_t$, then 

\begin{equation*}
\chi_{ab}=\frac{1}{2N}\Bigl(\dot h_{ab}-D_aN_b-D_bN_a\Bigr), \hskip 20pt
E_a=\frac{1}{N}\Bigl(D_a(N\phi)-\dot A_a+N^bD_bA_a+A_bD_aN^b\Bigr);
\end{equation*}
while the magnetic field strength is simply $B_{ab}=D_aA_b-D_bA_a$. 

Since the field equations of the EccHM system are second order, in the 3+1 
form of the model, the basic (Lagrangian) field variables are the 
configuration variables $(h_{ab},\Phi,\phi,A_a)$ and the corresponding 
velocities, $(\dot h_{ab},\dot\Phi,\dot\phi,\dot A_a)$. Then the Lagrangian 
density for the Maxwell field in its 3+1 form, ${\cal L}_M=-\frac{1}{2}E_a
E_bh^{ab}+ \frac{1}{4}B_{ab}B_{cd}h^{ac}h^{bd}$, is a function of the Lagrangian 
variables $h_{ab}$, $\phi$, $A_a$ and $\dot A_a$. 

However, the most convenient 3+1 form of the Lagrangian for the Higgs and 
gravitational sector of the model is also a first order one. This could be 
based on the decomposition \cite{IsNe} 

\begin{equation*}
R={\cal R}+\chi_{ab}\chi^{ab}-\chi^2+\frac{2}{N\sqrt{\vert h\vert}}\frac{\rm d}
{{\rm d}t}\bigl(\chi\sqrt{\vert h\vert}\bigr)+\frac{2}{N}D_a\bigl(D^aN-\chi 
N^a\bigr)
\end{equation*}
of the spacetime curvature scalar, where ${\cal R}$ is the curvature scalar 
of the spatial geometry $(\Sigma_t,h_{ab})$ and $h$ is the determinant of 
$h_{ab}$. By means of this decomposition we can write 

\begin{eqnarray*}
R\vert\Phi\vert^2\!\!\!\!&=\!\!\!\!&\bigl({\cal R}+\chi_{ab}\chi^{ab}-\chi^2
 \bigr)\vert\Phi\vert^2-\frac{2}{N}\chi\bigl(\dot\Phi\bar\Phi+\Phi\dot{\bar
 \Phi}\bigr)-\frac{2}{N}\bigl(D^aN-\chi N^a\bigr)D_a\vert\Phi\vert^2 \\
\!\!\!\!&+\!\!\!\!&\frac{2}{N\sqrt{\vert h\vert}}\frac{\rm d}{{\rm d}t}\bigl(
 \chi\vert\Phi\vert^2\sqrt{\vert h\vert}\bigr)+\frac{2}{N}D_a\bigl((D^aN-\chi 
 N^a)\vert\Phi\vert^2\bigr). 
\end{eqnarray*}
Thus, in the 3+1 form of the Higgs Lagrangian, it seems natural to drop the 
last two terms, which, after integration on $\Sigma_t$, would give a total 
time derivative and the integral of a total spatial divergence, respectively. 
Hence, we choose the 3+1 form of the Higgs Lagrangian to be 

\begin{eqnarray}
\hat{\cal L}_H\!\!\!\!&{}\!\!\!\!&:=\frac{1}{2}t^a\bigl(\nabbla_a\Phi\bigr)t^b
 \bigl(\nabbla_b\bar\Phi\bigr)+\frac{1}{2}h^{ab}\bigl(D_a\Phi+{\rm i}A_a\Phi
 \bigr)\bigl(D_b\bar\Phi-{\rm i}A_b\bar\Phi\bigr)-\frac{1}{2}\mu^2\vert\Phi
 \vert^2-\frac{1}{4}\lambda\vert\Phi\vert^4- \nonumber \\
-\!\!\!\!&{}\!\!\!\!&\frac{1}{12}\bigl({\cal R}+\chi_{ab}\chi^{ab}-\chi^2\bigr)
 \vert\Phi\vert^2+\frac{1}{6}\frac{1}{N}\chi\bigl(\dot\Phi\bar\Phi+\Phi
 \dot{\bar\Phi}\bigr)+\frac{1}{6}\frac{1}{N}\bigl(D^aN-\chi N^a\bigr)D_a
 \vert\Phi\vert^2, \label{eq:2.10}
\end{eqnarray}
where, by the definition of the time derivative, 

\begin{equation*}
t^a\nabbla_a\Phi=\frac{1}{N}\bigl(\dot\Phi+{\rm i}N\phi\Phi-N^aD_a\Phi\bigr). 
\end{equation*}
Forming the `mechanical Lagrangian' for the matter sector of the model, 
$\hat L:=\int_\Sigma({\cal L}_M+\hat{\cal L}_H)N{\rm d}\Sigma$, its formal, 
non-trivial variational derivatives with respect to the velocities are 

\begin{equation*}
\frac{\delta\hat L}{\delta\dot A_a}=E^a\sqrt{\vert h\vert}, \hskip 20pt
\frac{\delta\hat L}{\delta\dot{\bar\Phi}}=\frac{1}{2}\bigl(t^a\nabbla_a
\Phi+\frac{1}{3}\chi\Phi\bigr)\sqrt{\vert h\vert}=:\frac{1}{2}\Pi\sqrt{\vert 
h\vert}.
\end{equation*}
Thus, the canonical momentum conjugate to $\bar\Phi$ is $\frac{1}{2}\Pi$. 
The 3+1 form of the matter field equations, (\ref{eq:2.2a}) and 
(\ref{eq:2.4}), can also be recovered as the `mechanical' Euler--Lagrange 
equations for $\phi$, $A_a$ and $\Phi$ with the Lagrangian $\hat L$. In 
particular, the only constraint in the matter sector, viz. the Gauss 
constraint $D_aE^a=4\pi J_at^a$, is just the Euler--Lagrange equation for 
$\phi$, where the current $J_a$ was defined in (\ref{eq:2.2a}). 

Similarly, the formal variational derivatives of $\hat L$ with respect to the 
lapse $N$ and the shift $N^a$ gives minus the energy density and momentum 
density, respectively, calculated from (\ref{eq:2.3}), up to the Gauss 
constraint:

\begin{eqnarray*}
&{}&\varepsilon:=T_{ab}t^at^b=-\frac{1}{\sqrt{\vert h\vert}}\frac{\delta\hat L}
{\delta N}+\bigl(D_cE^c-4\pi J_ct^c)\phi, \\
&{}&\pi_a:=T_{cd}P^c_at^d=-\frac{1}{\sqrt{\vert h\vert}}\frac{\delta\hat L}
{\delta N^a}+\bigl(D_cE^c-4\pi J_ct^c)A_a. 
\end{eqnarray*}
Finally, by the Hamiltonian and momentum constraints of general 
relativity, 

\begin{equation}
\frac{1}{2}\bigl({\cal R}+\chi^2-\chi_{ab}\chi^{ab}\bigr)=\kappa\varepsilon
+\Lambda, \hskip 20pt
D_b\bigl(\chi^b{}_a-\chi\delta^b_a\bigr)=\kappa\pi_a, \label{eq:2.11}
\end{equation}
respectively, these expressions for the energy and momentum densities 
reproduce the ones calculated directly from (\ref{eq:2.5}). The advantage of 
this form of the energy and momentum densities is that these are expressions 
of the \emph{initial data} on the hypersurface $\Sigma_t$ for the evolution 
equations, rather than the corresponding \emph{solution} of the evolution 
equations. Note also that, in the derivation of the form of the energy (and, 
in general, also the momentum) density that we will use to define the 
instantaneous vacuum states, we used only the \emph{constraint}, but 
\emph{not} any of the evolution equations. This is needed to be consistent 
with the definition of the instantaneous vacuum states, which is based on 
the use of the \emph{constraint} equations only (see subsection 
\ref{sub-3.3}). 


\section{The instantaneous vacuum states with FRW 
symmetries}
\label{sec:3}

Using the energy and momentum densities, calculated from (\ref{eq:2.5}), 
one can form the energy functional and determine those field configurations 
that are the \emph{critical points} of this functional. Remarkably enough, 
if, in addition, we require these configurations to solve \emph{all} the 
constraints of the theory (i.e. the Gauss constraint and the constraints 
(\ref{eq:2.11}) of the gravitational sector, too), then the resulting 
configuration is only slightly more general than that for the initial data 
set in the presence of FRW symmetries \cite{Sz16}. Thus, in the present 
paper, we concentrate only on the FRW case directly, without considering 
the rather lengthy and laborious general analysis.

\subsection{The FRW symmetric configurations}
\label{sub-3.1}

Let $\Sigma_t:=\{t={\rm const}\}$ be the foliation of the FRW symmetric 
spacetime by the transitivity surfaces of the isometries, where $t$ is the 
proper time coordinate along the integral curves of the future pointing unit 
normals of the hypersurfaces $\Sigma_t$ (see e.g. \cite{HE}). Thus the lapse 
is $N=1$. Let $S=S(t)$ be the (strictly positive) scale function for which 
the induced metric on $\Sigma_t$ is $h_{ab}=S^2{}_1h_{ab}$, where ${}_1h_{ab}$ 
is the standard negative definite metric on the unit 3-sphere, the flat 
3-space and the unit hyperboloidal 3-space, respectively, for $k=1,0,-1$. 
The extrinsic curvature of the hypersurfaces is $\chi_{ab}=(\dot S/S)h_{ab}$, 
where over-dot denotes derivative with respect to $t$, and hence its trace 
is $\chi=3\dot S/S$. (The shift is chosen to be zero.) The curvature scalar 
of the intrinsic Levi-Civita connection on the hypersurfaces is ${\cal R}=
6k/S^2$. In the initial value formulation of Einstein's theory the initial 
data are $h_{ab}$ and $\chi_{ab}$, and hence in the present case $S$ and 
$\dot S$, restricted by the constraint equations. For the metric with FRW 
symmetries Einstein's equations are well known \cite{HE} to reduce to 

\begin{equation}
3\bigl(\frac{\dot S}{S}\bigr)^2=\Lambda+\kappa\varepsilon-3\frac{k}{S^2}, 
\hskip 20pt
3\frac{\ddot S}{S}=\Lambda-\frac{1}{2}\kappa\bigl(\varepsilon+3P\bigr). 
\label{eq:3.1}
\end{equation}
The first of these equations is just the Hamiltonian constraint, while the 
second is the evolution equation. Here $P:=-\frac{1}{3}h^{ab}T_{ab}$ is the 
isotropic pressure. The momentum constraint is satisfied identically. 

If the fields of the matter sector of the EccHM system are required to be 
invariant under the isometries of the spacetime, then all the fields with 
a spatial vector index must be vanishing and the Higgs field must be 
constant on the hypersurfaces $\Sigma_t$. Thus, the EccHM system restricted 
by the FRW symmetries reduces to the Einstein--conformally coupled Higgs 
(EccH) system with $D_a\Phi=0$. Then the field equation for the Higgs field 
is 

\begin{equation}
\ddot\Phi+3\frac{\dot S}{S}\dot\Phi=-\bigl(\mu^2+\frac{2}{3}\Lambda\bigr)
\Phi-\bigl(\lambda+\frac{1}{6}\kappa\mu^2\bigr)\bar\Phi\Phi^2.  
\label{eq:3.2}
\end{equation}
The initial data for the evolution equations is the quadruplet $(\Phi,S;\dot
\Phi,\dot S)$, or, equivalently, $(\Phi,S;\Pi,\chi)$, subject to the 
constraint part of (\ref{eq:3.1}).

\subsection{The energy density}
\label{sub-3.2}

\begin{figure}[htbp]
\centerline{\includegraphics[width=0.9\textwidth]{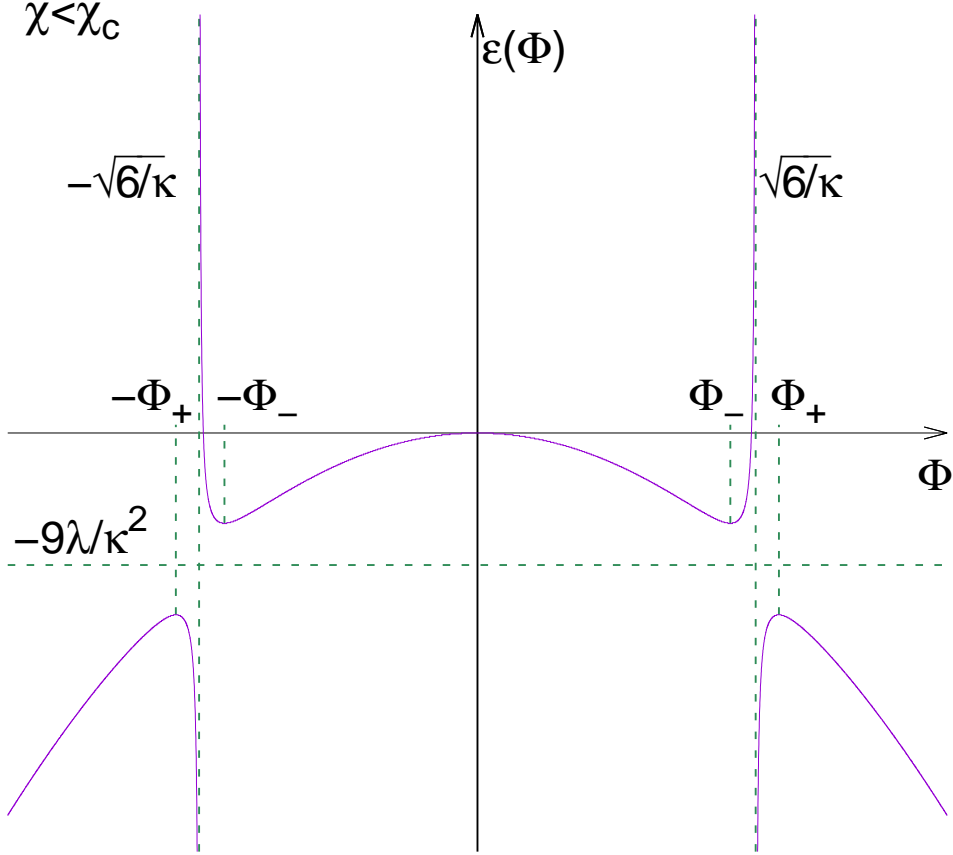}}
\caption{\label{fig:1}
The energy density $\varepsilon$ as a function of the real Higgs field $\Phi$ 
with $\Pi=0$ and given $\chi^2<\chi^2_c$. $\varepsilon(\Phi)$ has the `wine 
bottle' (rather than the familiar `Mexican hat') shape, in particular it has 
minima at $\pm\Phi_-$, only in the domain $\Phi^2<6/\kappa$. The critical 
points $\pm\Phi_+$ are maxima of $\varepsilon$. If $\chi\rightarrow\chi_c$, 
then $\Phi_{\pm}\rightarrow\sqrt{6/\kappa}$ and $\varepsilon(\Phi_{\pm})
\rightarrow-9\lambda/\kappa^2$. $\varepsilon(\Phi)$ is singular at $\Phi=
\pm\sqrt{6/\kappa}$.}
\end{figure}

Taking into account $\phi=0$, $A_a=0$, $D_a\Phi=0$ and the definition of 
$\Pi$, in the variables $(\Phi,S;\Pi,\chi)$ the energy density takes the form 

\begin{equation}
\varepsilon=\frac{1}{2}\frac{1}{1-\frac{1}{6}\kappa\vert\Phi\vert^2}\Bigl(
\vert\Pi\vert^2+\bigl(\mu^2+\frac{1}{3}\Lambda-\frac{1}{9}\chi^2\bigr)\vert
\Phi\vert^2+\frac{1}{2}\lambda\vert\Phi\vert^4\Bigr), \label{eq:3.3}
\end{equation}
the momentum density is zero, and the spatial stress is a pure trace, while 
the isotropic pressure is $P=\frac{1}{3}\varepsilon-\frac{1}{3}\mu^2\vert
\Phi\vert^2$. (\ref{eq:3.3}) shows that the energy density does not depend 
on the configuration variable $S$. For a given, fixed $\chi$ the function 
$\varepsilon(\Phi,\Pi,\chi)$ can have local minima precisely when $\chi^2<
\chi^2_c:=9(\mu^2+\Lambda/3+3\lambda/\kappa)$. They are at $\Pi=0$ and 
$\Phi$ solving 

\begin{equation}
\Bigl(1-\frac{1}{12}\kappa\vert\Phi\vert^2\Bigr)\lambda\vert\Phi\vert^2+
\bigl(\mu^2+\frac{1}{3}\Lambda-\frac{1}{9}\chi^2\bigr)=0, \label{eq:3.4}
\end{equation}
i.e. at $\Phi_v$ for which 

\begin{equation}
\vert\Phi_v\vert^2=\frac{6}{\kappa}\Bigl(1-\sqrt{1+\frac{\kappa}{3\lambda}
\bigl(\mu^2+\frac{1}{3}\Lambda-\frac{1}{9}\chi^2\bigr)}\Bigr). \label{eq:3.5}
\end{equation}
For given $\chi$ the graph of $\varepsilon=\varepsilon(\Phi,0,\chi)$ consists 
of two disconnected pieces, and it is only the domain $\vert\Phi\vert^2<6/
\kappa$, $\chi^2<\chi^2_c$ where it has the `wine bottle' (rather than the 
familiar `Mexican hat') shape. $\varepsilon(\Phi,0,\chi)$ is singular at 
$\vert\Phi\vert=\sqrt{6/\kappa}$. The energy density in the state $\Phi_v$ is 
$\varepsilon_v(\chi)=-\frac{1}{4}\lambda\vert\Phi_v\vert^4<0$. (For the energy 
density of the real Higgs field with $\mathbb{Z}_2:\Phi\mapsto-\Phi$ gauge 
symmetry, see Fig.~\ref{fig:1}.)

\begin{figure}[htbp]
\centerline{\includegraphics[width=0.9\textwidth]{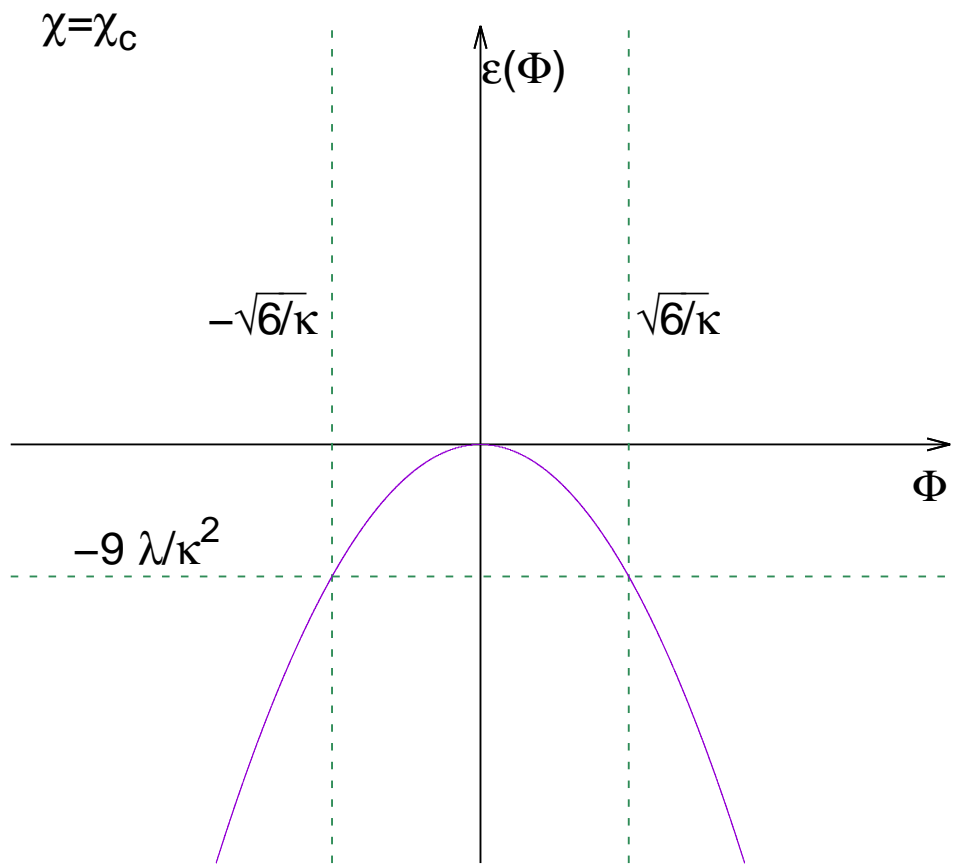}}
\caption{\label{fig:2}
The energy density $\varepsilon$ as a function of the real Higgs field $\Phi$ 
with $\Pi=0$ and $\chi^2=\chi^2_c$. $\varepsilon(\Phi)$ is not bounded from 
below.}
\end{figure}

\begin{figure}[htb]
\centerline{\includegraphics[width=0.9\textwidth]{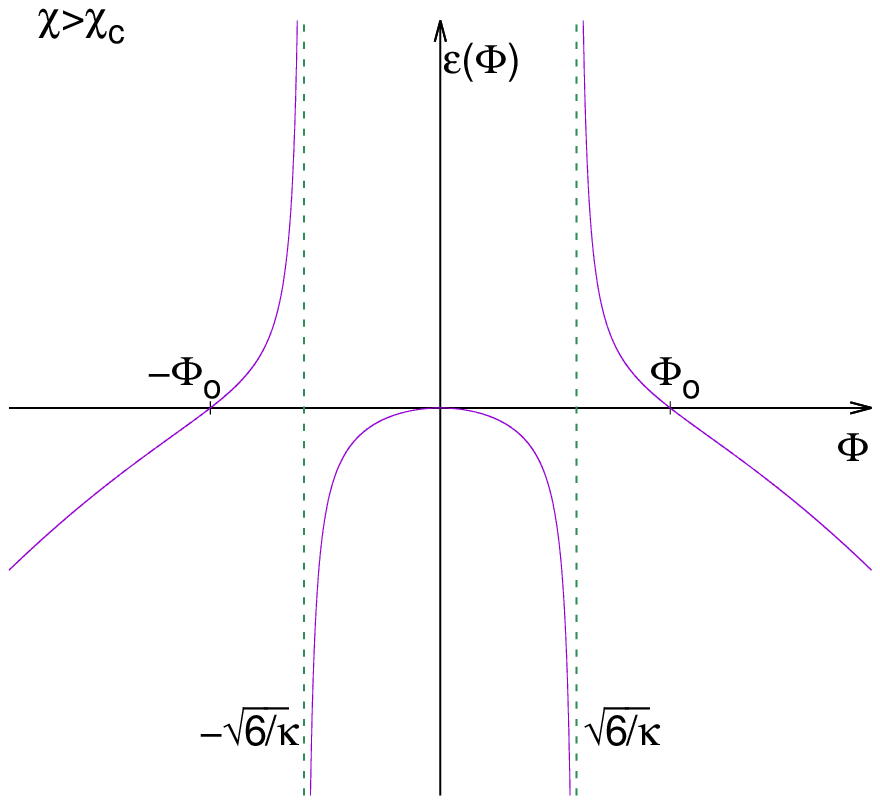}}
\caption{\label{fig:3}
The energy density $\varepsilon$ as a function of the real Higgs field $\Phi$ 
with $\Pi=0$ and given $\chi^2>\chi^2_c$. $\varepsilon(\Phi)$ is not bounded 
from below. If $\chi\rightarrow\chi_c$, then the zero $\Phi_0$ tends to 
$\sqrt{6/\kappa}$. $\varepsilon(\Phi)$ is singular at $\Phi=\pm\sqrt{6/
\kappa}$.}
\end{figure}

If $\chi\geq\chi_c$, then $\varepsilon(\Phi,0,\chi)$, as a function of 
$\Phi$, is \emph{not} bounded from below (Fig.~\ref{fig:2}, 
Fig.~\ref{fig:3}). (For a more detailed discussion of the function 
$\varepsilon=\varepsilon(\Phi,\Pi,\chi)$, see \cite{Sz16}.) 

\subsection{The instantaneous vacuum states}
\label{sub-3.3}

An instantaneous state of the physical system, represented by tensor fields 
on a spacelike hypersurface $\Sigma$ defining the `instant', will be called 
an \emph{instantaneous vacuum state} if the matter fields admit the isometries 
of the spacetime as symmetries, solve \emph{all the constraint} parts of the 
field equations and minimize the energy functional. Thus, in particular, the 
instantaneous vacuum states are required to be \emph{physical states}, i.e. 
points of the constraint surface in the phase space of the coupled 
matter+gravity system. Next we determine these states on the transitivity 
hypersurfaces in the FRW cosmological spacetimes. (For a more detailed 
discussion of this concept of vacuum state in classical field theory, as 
well as its particular form in the presence of the Kantowski--Sachs 
symmetries, representing e.g. the spacetime geometry inside spherical black 
holes, see \cite{Sz16}.) 

By (\ref{eq:3.5}), on a given transitivity hypersurface $\Sigma_t$ of the FRW 
symmetries the energy density can have local minima precisely when the mean 
curvature of $\Sigma_t$ satisfies the inequality $\chi^2<\chi^2_c$. Thus, 
instantaneous vacuum states cannot exist on hypersurfaces whose mean curvature 
is the critical value $\chi_c$ or higher. Since near the initial singularity 
of spacetime the mean curvature of the foliation $\Sigma_t$ diverges, 

\begin{equation}
\chi^2<\chi^2_c:=9\bigl(\mu^2+\frac{\Lambda}{3}+3\frac{\lambda}{\kappa}\bigr)
\simeq4.9\times10^{64}cm^{-2} \label{eq:3.6}
\end{equation}
is a non-trivial necessary condition for the existence of an instantaneous 
vacuum state. (In particular, for $\chi^2>\chi^2_c$ the expression under 
the square root in (\ref{eq:3.5}) would be negative. The numerical value of 
$\chi^2_c$ in (\ref{eq:3.6}) is given in the $\hbar=c=1$ units.) If such a 
vacuum state exists, then it is necessarily \emph{gauge symmetry breaking} 
(see Fig.~\ref{fig:1}), and depends on $\chi$. Indeed, $\vert\Phi_v\vert$ is 
\emph{completely determined} by $\chi$, but the $U(1)$ gauge transformation 
$\Phi_v\mapsto\exp({\rm i}\alpha)\Phi_v$, $\alpha\in[0,2\pi)$, takes an 
instantaneous vacuum state into a different such state. Therefore, the 
instant when the mean curvature is just the critical value $\chi_c$ will be 
the instant of the `genesis'/`evanescence' of rest masses, just when the 
BEH mechanism starts/ends to work (depending on whether the mean curvature 
is decreasing or increasing, respectively). As we will see, the time 
dependence of $\chi$ yields time dependence of the rest masses obtained via 
the BEH mechanism. 

Since the instantaneous vacuum states are defined to be certain \emph{physical 
states}, the corresponding field configuration must solve the Hamiltonian 
constraint. (The Gauss and the momentum constraints are satisfied 
identically.) If ${\cal R}_v$ denotes the curvature scalar of the intrinsic 
spatial geometry of the hypersurface $\Sigma_t$ in the vacuum state, then, 
by (\ref{eq:3.4}), it is 

\begin{equation}
\frac{1}{2}{\cal R}_v=\Lambda-\frac{1}{4}\kappa\lambda\vert\Phi_v\vert^4-
\frac{1}{3}\chi^2=\frac{1}{1-\frac{1}{12}\kappa\vert\Phi_v\vert^2}\bigl(
\Lambda+\frac{1}{4}\kappa\mu^2\vert\Phi_v\vert^2-\frac{1}{3}\chi^2\bigr), 
\label{eq:3.7}
\end{equation}
which is constant on $\Sigma_t$. By (\ref{eq:3.5}) the first two terms 
together in the brackets on the right is negative for any $\chi^2\leq\chi^2
_c$, and hence ${\cal R}_v$ must be \emph{negative}. Since ${\cal R}=6k/S^2$, 
this implies that the discrete parameter $k$ must be $-1$, and that the value 
$S_v$ of the scale function $S$ is also determined completely by $\chi$. 
Therefore, in particular, in \emph{globally defined} instantaneous vacuum 
states on $\Sigma_t$ which would also be invariant with respect to the 
isometries the intrinsic geometry of $\Sigma_t$ is hyperbolic: $k=-1$. In 
particular, the existence of such states requires that topologically 
$\Sigma_t$ be $\mathbb{R}^3$. 

On the other hand, if these states were required to be defined only on 
\emph{proper open subsets} of $\Sigma_t$ and their invariance under the whole 
isometry group were not required but were allowed to be $O(1,3)$-invariant 
even in the $k=1,0$ cases, then the existence of these states would not imply 
$k=-1$. These \emph{quasi-locally} defined instantaneous (gauge symmetry 
breaking) vacuum states would be enough to be able to define rest masses 
quasi-locally, and the non-zero rest masses could in fact be introduced via 
the BEH mechanism (see also subsections \ref{sub-1.1} and \ref{sub-1.1*}). 

Finally, it should be noted that the 1-parameter family of instantaneous 
vacuum states, parametrized by the mean curvature $\chi$, does \emph{not} 
solve the evolution equations, the second of (\ref{eq:3.1}) and 
(\ref{eq:3.2}): Substituting (\ref{eq:3.5}) into the evolution equations a 
tedious but straightforward calculation yields the contradiction $\Phi_v=0$. 
Therefore, the evolution equations take an instantaneous vacuum state into 
a non-vacuum state in the next instant.


\section{The genesis of the rest masses}
\label{sec:4}

\subsection{The strategy of the calculation of the rest masses}
\label{sub-4.1}

Our calculation of the rest masses via the BEH mechanism is based on 
the use of the instantaneous vacuum states. Thus we should assume that the 
spacetime satisfies those conditions that ensure the existence of the 
instantaneous vacuum states. In particular, the spacetime should admit a 
foliation by Cauchy surfaces $\Sigma_t$ with constant mean curvature, and 
that the mean curvature can be used as an extrinsic time coordinate (the 
`York time'), by means of which the hypersurfaces can also be labeled. 
Nevertheless, \emph{no} evolution equation will be used in these calculations. 

To ensure the existence of instantaneous vacuum states, condition 
(\ref{eq:3.6}) is assumed to be satisfied. Then the instantaneous vacuum 
states are necessarily \emph{gauge symmetry breaking}, and hence the BEH 
mechanism works. On the other hand, since the rest masses can be introduced 
even \emph{quasi-locally} (see subsection \ref{sub-1.1}), the instantaneous 
vacuum states are \emph{not} required to be global and the BEH mechanism 
still works.

\subsection{The BEH mechanism in the EccHM system}
\label{sub-4.2}

Suppose that the leaves $\Sigma_t$ of the foliation are of constant mean 
curvature. Let us choose the vacuum state, $\Phi_v$, to be \emph{real}. 
Then by an appropriate gauge transformation any Higgs field can be 
transformed into the form $\Phi=\Phi_v+H$, where $H$ is a \emph{real} 
function. In fact, the existence of such a gauge (the so-called `unitary 
gauge') is a consequence of a much more general result, proven by Weinberg, 
even for arbitrary Higgs multiplet and any \emph{compact} gauge group 
\cite{We73}. Then, in this gauge, the matter sector of the instantaneous 
vacuum states is characterized by $\phi=0$, $A_a=0$, $H=0$ and $E_a=0$, 
$\dot H=0$, $\Pi=0$. The latter implies that $t^a\nabla_a\Phi_v=-\frac{1}{3}
\chi\Phi_v$. Note that, in the gravitational sector of the instantaneous 
vacuum states, the extrinsic curvature is a pure trace: $\chi_{ab}=\frac{1}{3}
\chi h_{ab}$. 

Rewriting the Lagrangian density $\hat{\cal L}:=\hat{\cal L}_H+{\cal L}_M$ 
in terms of the variables $\phi$, $A_a$, $H$ and their time derivative, its 
derivative with respect to the gauge potentials (while keeping their 
derivatives fixed) at the instantaneous vacuum state are 

\begin{eqnarray*}
&{}&\bigl(\frac{\partial\hat{\cal L}}{\partial\phi}\bigr)_v=\bigl(
 \frac{\partial\hat{\cal L}_H}{\partial\phi}\bigr)_v=\frac{\rm i}{2}\Bigl(
 \Phi t^c\nabbla_c\bar\Phi-\bar\Phi t^c\nabbla_c\Phi\Bigr)_v=0, \\
&{}&\bigl(\frac{\partial\hat{\cal L}}{\partial A_a}\bigr)_v=\bigl(
  \frac{\partial\hat{\cal L}_H}{\partial A_a}\bigr)_v=\frac{\rm i}{2}\Bigl(
 \Phi\mathbb{D}^a\bar\Phi-\bar\Phi\mathbb{D}^a\Phi\Bigr)_v=0. 
\end{eqnarray*}
Here, in the first equation we used that in the vacuum state $E_a=0$, $D_a
\Phi_v=0$ and $\Pi_v=0$; while in the second that $A_a=0$ and $D_a\Phi_v=0$. 
These two equations can be summarized as $(\partial\hat{\cal L}/\partial
\omega_a)_v=((\partial\hat{\cal L}/\partial\phi)t^a+(\partial\hat{\cal L}/
\partial A_b)P^a_b)_v=0$. The first derivative of $\hat{\cal L}$ with respect 
to $\Phi$ (while keeping $\dot\Phi$ and $D_a\Phi$ fixed) at the instantaneous 
vacuum state is 

\begin{eqnarray*}
\bigl(\frac{\partial\hat{\cal L}}{\partial\Phi}\bigr)_v\!\!\!\!&=\!\!\!\!&
 -\frac{1}{2}\mu^2\Phi_v-\frac{1}{2}\lambda\Phi^3_v-\frac{1}{12}\bigl({\cal 
 R}_v+\chi_{ab}\chi^{ab}-\chi^2\bigr)\Phi_v-\frac{1}{18}\chi^2\Phi_v= \\
\!\!\!\!&=\!\!\!\!&-\frac{1}{2}\Bigl(\frac{1}{6}\bigl({\cal R}_v+\chi^2-
 \chi_{ab}\chi^{ab}\bigr)+\mu^2+\lambda\vert\Phi_v\vert^2-\frac{1}{9}\chi^2\Bigr)
 \Phi_v= \\
\!\!\!\!&=\!\!\!\!&-\frac{1}{2}\Bigl(-\frac{1}{12}\kappa\lambda\vert\Phi_v
 \vert^4+\lambda\vert\Phi_v\vert^2+\mu^2+\frac{1}{3}\Lambda-\frac{1}{9}\chi^2
 \Bigr)\Phi_v=0.
\end{eqnarray*}
Here, in the second step we used $\chi_{ab}=\chi h_{ab}/3$, in the third step 
the Hamiltonian constraint and the expression of the energy density 
$\varepsilon_v(\chi)$ in the vacuum state (see subsection \ref{sub-3.2}), 
and in the last step (\ref{eq:3.4}). This yields 

\begin{equation*}
\bigl(\frac{\partial\hat{\cal L}}{\partial H}\bigr)_v=\Bigl(\frac{\partial
\hat{\cal L}}{\partial\Phi}+\frac{\partial\hat{\cal L}}{\partial\bar\Phi}
\Bigr)_v=0.
\end{equation*}
Therefore, \emph{the instantaneous vacuum state is a critical point} of the 
Lagrangian $\hat{\cal L}$ both with respect to the gauge and the Higgs fields, 
and hence their rest mass is well defined. In fact, the rest mass for the 
gauge field, 

\begin{equation}
m^2_\omega:=\frac{1}{4}\bigl(g_{ab}\frac{\partial^2\hat{\cal L}}{\partial\omega_a
\partial\omega_b}\bigr)_v=\frac{1}{4}\bigl(g_{ab}\frac{\partial^2\hat{\cal L}
_H}{\partial\omega_a\partial\omega_b}\bigr)_v=\Phi_v^2, 
\label{eq:4.1}
\end{equation}
is well defined and \emph{positive}\footnote{In the particle physics 
literature, instead of the 4-covariant connection 1-form $\omega_a$ the 
4-potential $\varpi_a:=\omega_a/g$ is used, where $g>0$ is the coupling 
constant; and the corresponding rest mass is defined by the second 
derivative of the Lagrangian with respect to $\varpi_a$ rather than to 
$\omega_a$. With this convention $m_\varpi=g\vert\Phi_v\vert$.}. 
To calculate the rest mass for $H$, first we compute the second derivatives 
of $\hat{\cal L}_H$ with respect to $\Phi$ and $\bar\Phi$: 

\begin{eqnarray*}
\bigl(\frac{\partial^2\hat{\cal L}_H}{\partial\Phi^2}\bigr)_v\!\!\!\!&=
 \!\!\!\!&-\frac{1}{2}\lambda\Phi^2_v, \\
\bigl(\frac{\partial^2\hat{\cal L}_H}{\partial\Phi\partial\bar\Phi}\bigr)_v
\!\!\!\!&= \!\!\!\!&-\frac{1}{2}\mu^2-\lambda\Phi_v^2-\frac{1}{12}
 \bigl({\cal R}_v+\chi^2-\chi_{ab}\chi^{ab}\bigr)+\frac{1}{9}\chi^2= \\
\!\!\!\!&=\!\!\!\!&-\frac{1}{2}\Bigl(-\frac{1}{12}\kappa\lambda\Phi_v^4+2
 \lambda\Phi_v^2+\mu^2+\frac{1}{3}\Lambda-\frac{2}{9}\chi^2\Bigr)=-\frac{1}
 {2}\bigl(\lambda\Phi_v^2-\frac{1}{9}\chi^2\bigr). 
\end{eqnarray*}
Thus, finally, the rest mass of the field $H$ is 

\begin{equation}
m^2_H:=-\bigl(\frac{\partial^2\hat{\cal L}}{\partial H^2}\bigr)_v=-\Bigl(
\frac{\partial^2\hat{\cal L}_H}{\partial\Phi^2}+2\frac{\partial^2\hat{\cal 
L}_H}{\partial\Phi\partial\bar\Phi}+\frac{\partial^2\hat{\cal L}_H}{\partial
\bar\Phi^2}\Bigr)_v=2\lambda\Phi_v^2-\frac{1}{9}\chi^2. 
\label{eq:4.2}
\end{equation}
Since, to guarantee the existence of instantaneous vacuum states we assumed 
that $\chi^2<\chi^2_c$, by (\ref{eq:3.5}) the rest mass of the field $H$ is 
\emph{positive}.

\subsection{The time dependence of the rest masses}
\label{sub-4.3}

By (\ref{eq:3.5}) the norm $\vert\Phi_v\vert$ depends on the mean curvature, 
and hence the rest masses $m_\omega$ and $m_H$ are \emph{time dependent} if 
$\dot\chi\not=0$, though they have \emph{different} time dependence. In 
particular, both are monotonically \emph{decreasing} with decreasing 
$\chi^2$. At the instant of their `genesis', i.e. in the $\chi^2\rightarrow
\chi^2_c$ limit, $m^2_\omega\rightarrow3/\kappa$ and $m^2_H\rightarrow9
\lambda/\kappa-\mu^2-\Lambda/3\simeq1.7\times10^{64}cm^{-2}$ (in the $\hbar=
c=1$ units). 

Since $({\rm d}\vert\Phi_v\vert^2/{\rm d}\chi^2)$ diverges if $\chi^2
\rightarrow\chi^2_c$ and $({\rm d}\vert\Phi_v\vert^2/{\rm d}\chi^2)
\rightarrow0$ if $\chi\rightarrow0$, the time dependence of the rest masses 
is significant only just after their `genesis'. For example, while the Hubble 
time corresponding to $\chi_c$ is $t_c:=3/\chi_c\simeq 4.5\times10^{-43}sec$ 
(which is almost ten Planck times), the Higgs mass decreased to the 
half of its initial value (at the instant of its `genesis') by $5.8\times10
^{-43}sec$ Hubble time (i.e. c.c. in the next three Planck times); and it 
decreased to twice of its present value, viz. to $2\times(6.2\times10^{15}cm
^{-1})$, by $5.5\times10^{-26}sec$ Hubble time. Remarkably enough, the 
characteristic time of the weak interactions that the Higgs mass parameter 
defines is $1/c\vert\mu\vert\simeq5.4\times10^{-27}sec$. Hence, at this 
characteristic time, the rest mass of the Higgs field was roughly twice of 
its present value. 

On the other hand, for small enough $\chi^2$ the norm $\vert\Phi_v\vert$ can 
be expanded as 

\begin{equation}
\vert\Phi_v\vert^2=\frac{6}{\kappa}\Bigl(1-\sqrt{1+\frac{\kappa}{3\lambda}
(\mu^2+\frac{1}{3}\Lambda-\frac{1}{9}\chi^2)}\Bigr)=-\frac{\mu^2}{\lambda}+
\frac{\kappa}{12}\frac{\mu^4}{\lambda^2}-\frac{1}{3\lambda}\bigl(\Lambda-
\frac{1}{3}\chi^2\bigr)+.... \label{eq:4.4}
\end{equation}
The first term in the expansion is the well known vacuum value in the 
Standard Model in Minkowski spacetime (see e.g. \cite{AL}), the second, being 
proportional to Newton's gravitational constant, is of proper gravitational 
origin, while the third, containing the cosmological constant and the rate 
of expansion of the universe, has cosmological origin. In the $\chi^2
\rightarrow0$ limit $\vert\Phi_v\vert^2$ reduces to that obtained in 
subsection \ref{sub-2.2} for the norm of the Higgs field in the `spacetime 
vacuum state' (see equation (\ref{eq:2.9})). The present value of the Hubble 
constant in our observed Universe is $(\dot S/S)_{\rm now}=\frac{1}{3}
\chi_{\rm now}\simeq 7\times 10^{-28}cm^{-1}$. Hence, the corrections in 
(\ref{eq:4.4}) to the value $\sqrt{-\mu^2/\lambda}$ of the norm of the Higgs 
field in the symmetry breaking vacuum state in the Poincar\'e invariant 
Standard Model are extremely tiny, and they have significance only in the 
extreme gravitational circumstances. 


\section{Summary, conclusions and final remarks}
\label{sec:5}

We investigated certain \emph{kinematical} consequences of the conformally 
invariant coupling of the Higgs field to Einstein's theory of gravity. First, 
we showed that global \emph{spacetime} vacuum states, i.e. which would have 
maximal spacetime symmetry, solve the field equations and minimize the energy 
density, do \emph{not} exist. Then, we showed that in the $k=0,1$ FRW 
spacetimes \emph{global instantaneous} vacuum states, i.e. field 
configurations on the transitivity hypersurfaces of the spacetime symmetries 
which would be invariant with respect to these symmetries, solve the 
\emph{constraint} parts of the field equations and minimize the energy 
functional, do \emph{not} exist. Also, even general \emph{quasi-local} 
instantaneous vacuum states (i.e. which are represented by field 
configurations that are not necessarily globally defined on the spacelike 
hypersurfaces) do \emph{not} exist on hypersurfaces whose mean curvature is 
greater than a large, but finite critical value. If the mean curvature is 
less than this critical value, then instantaneous vacuum states exist, 
which are necessarily gauge symmetry breaking and depend on the mean 
curvature. 

Using this concept of the global or quasi-local instantaneous (gauge symmetry 
breaking) vacuum states, in spacetimes that admit a foliation by constant 
mean curvature Cauchy hypersurfaces and the mean curvature can be used as a 
time coordinate, we determined how the rest mass of the matter fields of the 
Einstein-conformally coupled Higgs-Maxwell system depends on the extrinsic 
York time parameter of the hypersurfaces. We found that there are extreme 
gravitational situations in which \emph{the notion of rest mass} of the 
Higgs field, zero or non-zero, \emph{cannot be introduced}. In these 
situations the Higgs field does not have particle interpretation. The 
resulting non-zero rest masses of the fields, introduced via the BEH 
mechanism, are time dependent in a non-stationary spacetime and they are 
\emph{decreasing} with decreasing mean curvature. 

Therefore, according to the present model, the scenario of the genesis of the 
observed non-zero rest mass of the fields is rather different from the usual 
view: According to the traditional picture initially, in the very early stage 
of the history of the Universe, the Higgs field was in the \emph{symmetric 
vacuum state} and the gauge and the fermion fields had \emph{zero rest mass}, 
and at a (mathematically still not yet specified) later stage the Higgs field 
developed into a \emph{symmetry braking vacuum} and the gauge and fermion 
fields got non-zero rest mass via the BEH mechanism. On the other hand, 
according to the present model, initially the Higgs field \emph{did not have 
any vacuum state} and \emph{its rest mass was not defined at all}, while the 
gauge field had zero rest mass. When the mean curvature decreased below the 
critical value, symmetry breaking vacuum states of the Higgs field emerged, 
and the gauge and the Higgs fields got enormously large rest masses. These 
rest masses were \emph{decreasing} rapidly to their present value with the 
expansion of the Universe. 

In fact, the key statements in the present simple model hold true in the 
more realistic Einstein-conformally coupled Standard Model (EccSM) system, 
in which the gauge group could be any compact Lie group, the Higgs field is 
a multiplet of complex self-interacting fields and there could be any 
collection of Weyl spinor fields coupled in the minimal way to the gauge 
fields and via the Yukawa coupling to the Higgs multiplet \cite{Sz16}. In 
particular, in the Einstein-conformally coupled Weinberg--Salam model the 
rest mass of the electron, and the W and Z bosons is $m_e=\frac{1}
{\sqrt{2}}G_e\vert\Phi_v\vert$, $m_W=\frac{1}{2}g\vert\Phi_v\vert$, $m_Z=
\frac{1}{2}\sqrt{g^2+g'{}^2}\vert\Phi_v\vert$, respectively. Here $G_e$ is 
the appropriate Yukawa coupling, and $g$ and $g'$ are the $SU(2)$ and $U(1)$ 
coupling constants, respectively. Thus all of these have the \emph{same} time 
dependence via $\vert\Phi_v\vert$, which is different from that of the Higgs 
rest mass given by (\ref{eq:4.2}) above. Since in the Weinberg--Salam model 
electromagnetism is an emergent phenomenon due to the $U(2)\rightarrow U(1)$ 
gauge reduction during the symmetry breaking, in the Einstein-conformally 
coupled Weinberg--Salam model electromagnetism and the electric charge, 
$\vert e\vert=gg'(g^2+g'{}^2)^{-1/2}$, emerge at the same instant when the 
rest masses, at $\chi=\chi_c$. On the other hand, contrary to the non-zero 
rest masses, the charge $e$ does \emph{not} depend on $\chi$. 

If the mean curvature of the hypersurfaces happened to be increasing and 
exceeded the critical value, then the massive fields would lose their rest 
mass. In our (asymptotically exponentially expanding) Universe the mean 
curvature asymptotically tends to the finite, constant value $\sqrt{\Lambda/
3}$. Hence the `reverse--BEH' mechanism at the cosmological scale cannot 
provide the way in which the fields lose their rest mass. The `reverse-BEH' 
mechanism takes place inside black holes, deeply behind the event horizon 
near the central singularity. In fact, the specific analysis of section 
\ref{sec:3} can be carried out in the presence of Kantowski--Sachs (rather 
than FRW) symmetries, describing the spacetime geometry inside spherical 
black holes, and one can determine the (quasi-local) instantaneous vacuum 
states and calculate the rest masses \cite{Sz16}. The matter fields falling 
into a spherical black hole loose their rest mass before hitting the central 
singularity. 

The time dependence (and the \emph{different} time dependence) of the Higgs 
and the other fields is significant only when the mean curvature is close 
to its critical value. Thus, it may have a potential significance in the 
particle physics processes in the very early Universe (or near the central 
singularity in spherical black holes). Hence it could be interesting to see 
whether or not this time dependence, and in particular the fact that at the 
characteristic time scale defined by the Higgs rest mass parameter the rest 
masses were twice their present value, could yield observable effect in the 
particle genesis era of the very early Universe.


\bigskip

The author is grateful to \'Arp\'ad Luk\'acs, P\'eter Vecserny\'es and 
Gy\"orgy Wolf for the numerous and enlightening discussions both on the 
structure of the Standard Model and on various aspects of the present 
suggestion. Special thanks to Gy\"orgy Wolf for the careful reading of 
an earlier version of the paper, his suggestions to improve the text at 
several points and for drawing the figures; and to Helmut Friedrich and 
Paul Tod for their remarks on both the conformal cyclic cosmological model 
and the present suggestions. Thanks are due to the `Geometry and Relativity' 
program at the Erwin Schr\"odinger International Institute for Mathematics 
and Physics, Vienna, for the support and hospitality where the final 
version of the present paper was prepared.


\end{document}